# Short and Long Ranged Impurities in Fractional Quantum Hall Systems

B. A. Friedman,[1]

[1]*Department of Physics, Sam Houston State University, Huntsville, 77381, USA*

Short and long-range impurities have been examined for fractional quantum Hall systems. There appears to be a consistent computational picture for short range impurities. In the case of long range impurities, calculations agree qualitatively with experiment, in that the critical mobility is very sensitive to long range impurities and the critical mobility for long range impurities is larger than the critical mobility for short range impurities. The physical mechanism of this sensitivity and a quantitative understanding remain a challenging computational issue.

## I. INTRODUCTION

In this paper, short and long ranged impurities are studied in fractional quantum Hall systems. Recall that a quantum Hall system is a two-dimensional system of electrons at low temperature under strong magnetic fields[1,2]. The filling fraction, the number of electrons per magnetic flux quantum, is a fraction, for example, for filling $1/3$ , there is one electron per three flux quanta. These systems are prototypical strongly correlated electron systems and exhibit novel physics, for example, fractionally charged quasi particles. The particular physical system of interest are GaAs hetero structures, the original and arguably best characterized system where fractional quantum Hall physics occurs.

The existence of a fractional quantum Hall state is critically dependent on the strength of disorder, i.e. fractional quantum Hall behavior was not observed before $\delta$ - doping had been invented and low disorder strength systems were available. For what value of the disorder strength and type of disorder can fractional quantum Hall behavior be observed? This issue is of applied importance since it has been proposed that fractional quantum Hall systems can be used for quantum computing[3]. For such a purpose, one needs a filling fraction with non-abelian quasi particles, the simplest possible example is the fractional quantum Hall system with filling fraction $5/2$ . The best theoretical candidate to describe this system is the Moore-Read state which indeed has non-abelian quasi particles[4]. The advantage of quantum computing based on a material with non-abelian quasi particles is that in principle error correction is not necessary, there is a topological "barrier" to the influence of the environment. (Strictly speaking, filling fraction $5/2$ is too "simple" to realize a complete set of topologically protected gates).

Our current interest about disorder was motivated by the recent experiments[5] by the Purdue group on the effect of controlled amounts of short range impurities on the $5/2$ fractional quantum Hall state. These experiments show that samples which have much lower mobility than previous samples still can have quantum Hall behavior if the impurities are short ranged. To reiterate, the value of the mobility where fractional quantum Hall behavior ceases, the critical mobility $\mu_c$ is much lower when the dominant source of scattering is from short range impurities.

Independent of these very interesting experimental developments, there was a computational study of the effect of disorder on the entanglement entropy of fractional quantum Hall systems. Using a model of short range disorder introduced in[6,7], the entanglement entropy for the $1/3$ and $5/2$ states was calculated as a function of the disorder strength[8]. The entanglement entropy for the purposes of the present paper (and[8]) is the Von Neumann entropy of the reduced density matrix when the whole system is partitioned into two parts; call them the subsystem and the environment.

This quantity measures the quantum entanglement of the subsystem and the environment and is familiar from quantum information considerations. The entanglement entropy has good formal properties[9], it obeys strong subadditivity, however, it is difficult to calculate, in comparison to other Renyi entropies[10] and it is very difficult to measure directly as it involves the reduced density matrix of a many body system. Direct diagonalization results[8], by necessity on small systems, indicated no obvious signature of a transition in the entanglement entropy for $1/3$ filling as a function of disorder strength. However, for the filling $5/2$ there is a plateau then a sharp drop, call the value where this drop takes place $W_c$ . Recent, very interesting work[11] using an effective interaction for $1/3$ filling indicate a similar drop in the entanglement entropy. (The effective interaction is the Haldane pseudo potential[11], which yields the Laughlin state as the exact ground state, the previous calculation used long range Coulomb interaction for electron-electron interaction, both calculations used white noise for the impurity potential.)

Taking the value of $W_c$ from[8] and using the Born approximation result of[6,7] one can convert $W_c$ to mobility. This value of the mobility, agrees very well, with no adjustable parameters, with $\mu_c$ obtained from experiment. More precisely, $\mu_c$ from experiment[5] is the mobility for which the transport gap for the quantum Hall state vanishes.

The earliest work[6,7], the source of the model for disorder in fractional quantum Hall systems, calculated for $1/3$ filling, the boundary average Hall conductance, the Chern number, from which the transport gap can be extracted. These calculations indicate to get good agreement with experiment, a parameter describing layer thickness must be introduced. One is faced with the apparently paradoxical situation where layer thickness is unnecessary to describe short range disorder in the $5/2$ state, while layer thickness is needed for $1/3$ filling.

To help resolve this situation in the second section $\mu_c$ for ${}^{1}/_{3}$ filling will be reexamined using another procedure introduced in[7]. This method, based on gaps between energy eigenvalues will also be used to reexamine $\mu_c$ for the ${}^{7}/_{3}$ state. In the third section, a model of long range disorder is introduced, in fact, a simple model for δ - doping and some preliminary results for the ${}^{5}/_{2}$ state are given. The last section are conclusions and prospects for further computational/theoretical work.

## II. SHORT RANGE DISORDER AT ${}^{1}/_{3}$ AND ${}^{7}/_{3}$ FILLING FRACTIONS

The model used for short range disorder, introduced in[6,7] consists of spin-polarized electrons[12,13] confined to a single[13] Landau level that interact with the Coulomb potential. The random potential $U(r)$ is taken to have zero mean, to be Gaussian, and delta correlated

$$< U(r)U(r') > = W^2\, \delta(r - r')\ , (1)$$

and is projected into the fractionally filled Landau level. The numerical method used is direct diagonalization applied to square (aspect ratio one) clusters with periodic boundary conditions, the square torus geometry. For our calculations, only the ground state, or low lying excited states are needed, hence an iterative method, i.e. the Lanczos and Davidson algorithms have been used. The size of the state space, for example, for 10 electrons in 30 orbitals

$$= \binom{30}{10} \approx 3\, X\, 10^7\ \ , (2)$$

somewhat large considering the amount of averaging that needed to be done. Note however, from a strict memory stand point, several more system sizes are accessible.

The method used to calculate the critical disorder strength $W_c$ is to plot the average gap between the second and third excited states (the third and fourth states) vs. disorder strength. The value of W where the curve has a minimum is taken to be $W_c$ . The choice of the second and third states is appropriate since filling $^1/_3$ is being considered. For a very large system, one would expect, the value of W where the average gap first vanishes is $W_c$ . Here and elsewhere in this paper average means an average over realizations of disorder. However, for a finite size system, numerical evidence suggests this takes place at the minimum value of the gap[7].

Figure 1

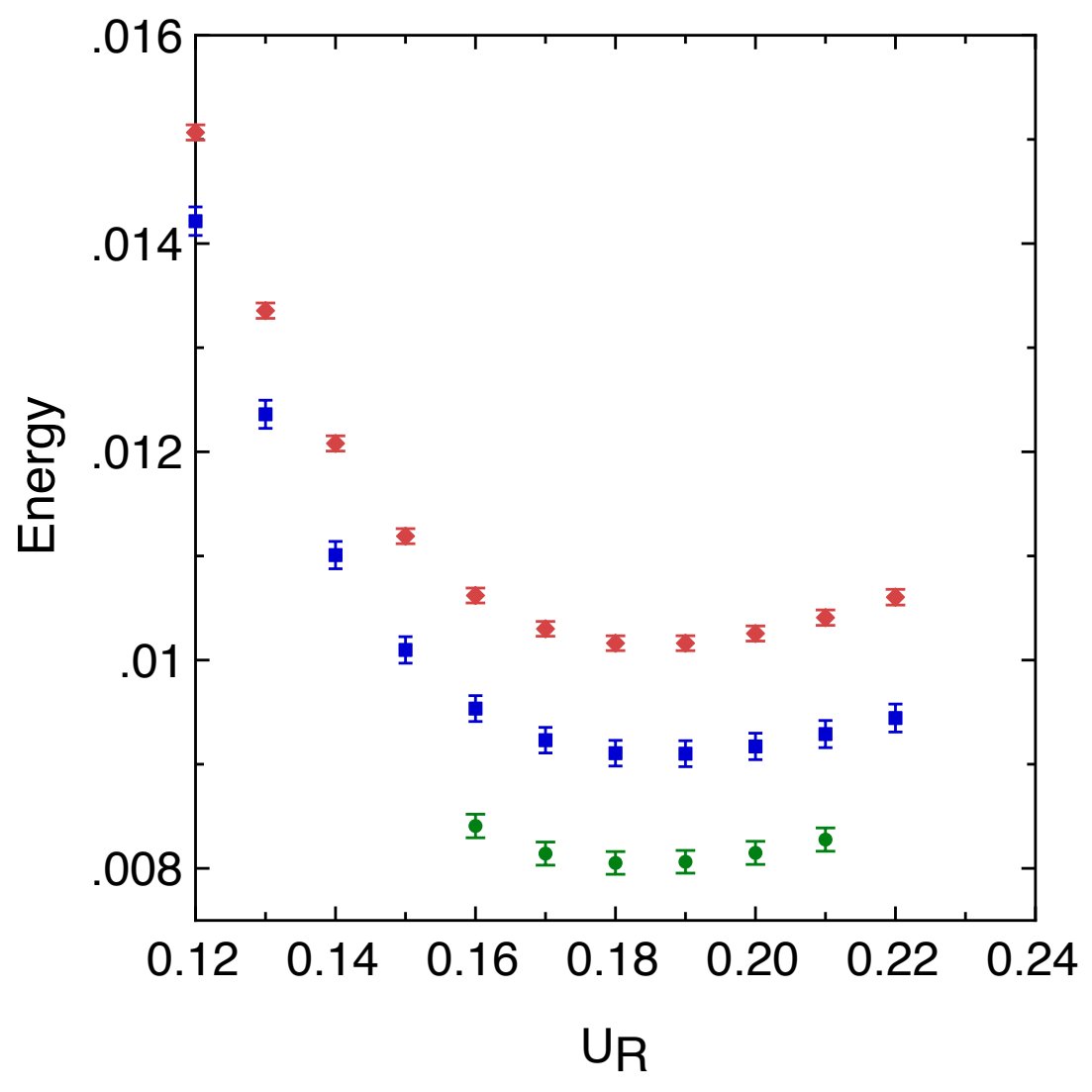

**FIG. 1. The average gap between the second and third excited states vs. $U_R$ ($U_R = \sqrt{^3/_2}\, W$) is plotted for $^1/_3$ filling in the lowest Landau level (n=0). The Davidson algorithm was used to calculate the first four eigenvalues, for system size 6 electrons in 18 orbitals (red diamonds), 7 electrons in 21 orbitals (blue squares) and 8 electrons in 24 orbitals (green circles).**

In figure 1, the average gap between the second and third excited states vs. $U_R$ ($U_R = \sqrt{^3/_2}\, W$) is plotted for $^1/_3$ filling in the lowest Landau level (n=0). The Davidson algorithm was used to calculate the first four eigenvalues, for system size 6 electrons in 18 orbitals, 7 electrons in 21 orbitals and 8 electrons in 24 orbitals. One sees a minimum at roughly $U_R \approx .185$ ( $W = .15$ ) and there is not much system size dependence. This translates via the Born approximation to a value of $\mu_c$ of $9.3\ X\ 10^4 \frac{cm^2}{Vs}$ . From[6,14], the measured value of $\mu_c$ is $1.0\ X\ 10^5 \frac{cm^2}{Vs}$ (This value of $\mu_c$ does not come from the recent experiments on controlled short range disorder. The experimental system of[5] could not access the physical conditions necessary to study disorder at $^1/_3$ filling[15]). One thus obtains very good agreement with experiment without the effect of finite layer thickness similar to the situation for filling $^5/_2$ .

Figure 2

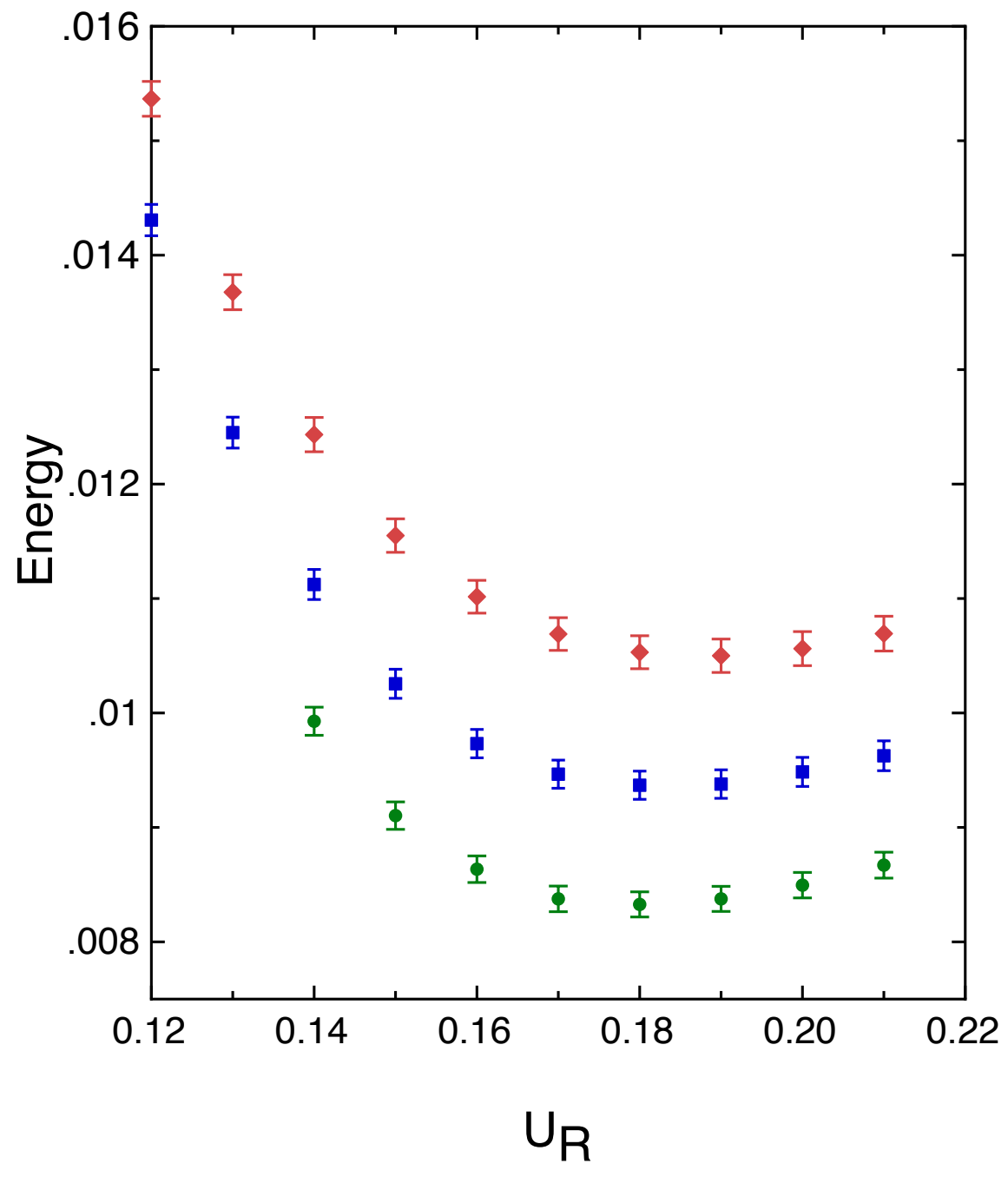

**FIG. 2. Average gap in between the third and fourth eigenvalues of the Lanczos matrix vs disorder $U_R$ for $^1/_3$ filling in the lowest Landau level (n=0). The red diamonds are for 6 electrons in 18 orbitals, the blue squares are for 7 electrons in 21 orbitals and the green circles are for 8 electrons in 24 orbitals.**

It is numerically useful to note that one can get $\mu_c$ by using the Lanczos algorithm to calculate the ground state. That is, the Lanczos algorithm gives a tridiagonal matrix which one truncates at a sufficiently large dimension to get the accurate ground state. One can use the average gap between the third and fourth eigenvalues of this matrix as a quantity to plot vs. disorder. In figure 2, this gap is plotted vs. disorder strength $U_R$ . The plot is close to figure 1, in particular the value of $U_R$ where the gap is a minimum is close to $U_R \approx .185$ .

Figure 3

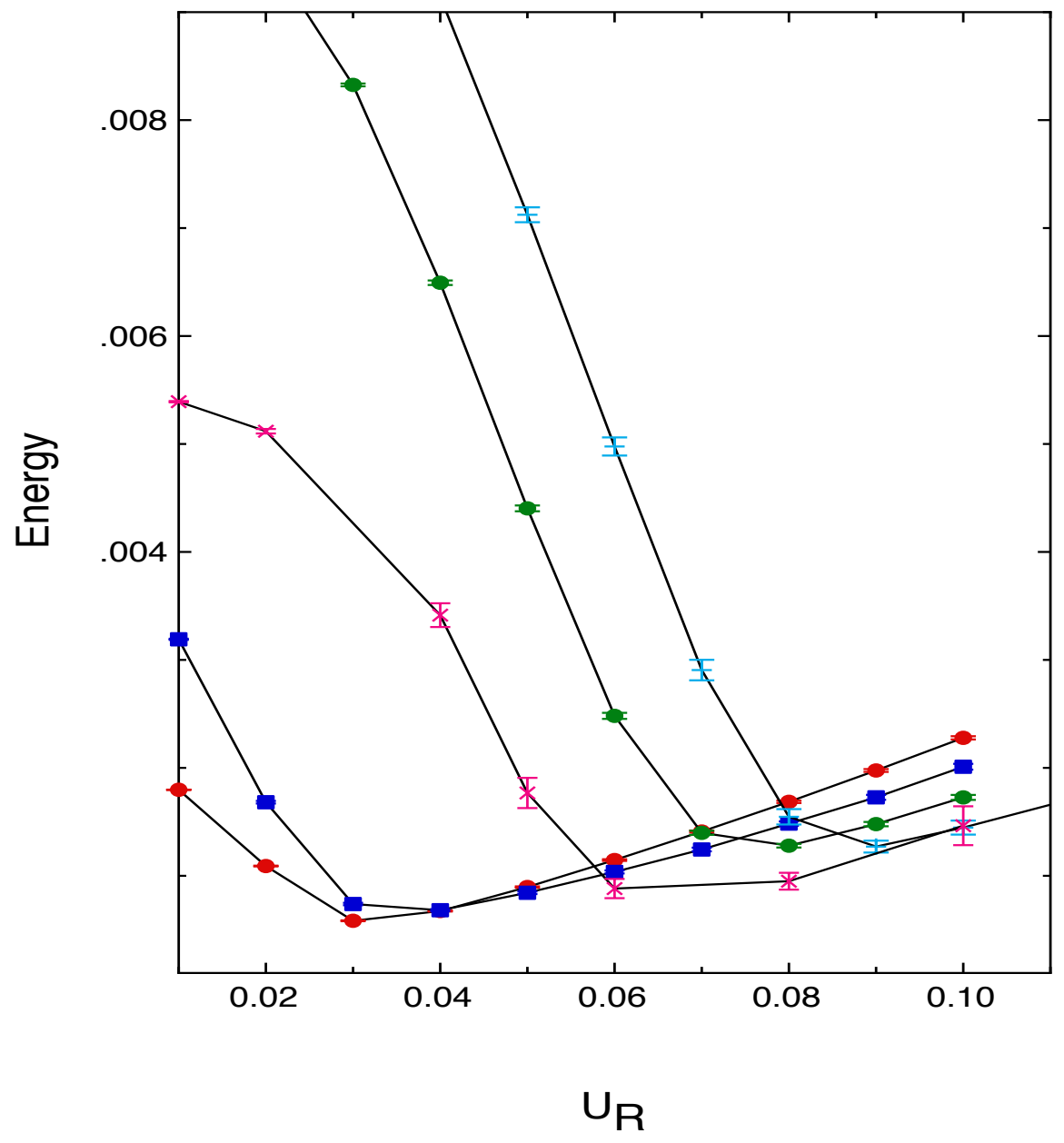

**FIG. 3. Average gap in between the third and fourth eigenvalues of the Lanczos matrix vs disorder $U_R$ for $^7/_3$ filling in the second Landau level (n=1). The red diamonds are for 6 electrons in 18 orbitals, the blue squares are for 7 electrons in 21 orbitals, the green circles are for 8 electrons in 24 orbitals, the blue crosses are for 9 electrons in 27 orbitals, and the pink Xes are for 10 electrons in 30 orbitals. The lines are a guide for the eye.**

Using the above observation, we turn to the second (n=1) Landau level. Experimental evidence suggests that filling $^7/_3$ is topologically equivalent to the Laughlin $^1/_3$ state[16], hence the gap between the third and fourth eigenvalues of the Lanczos matrix has been calculated and plotted vs.

disorder strength in figure 3. One notes there is considerable system size dependence, 6 electrons in 18 orbitals and 7 electrons in 21 orbitals form one group of curves, while 8-24, 9-27 and 10-30 are another group. This is similar behavior as to what is observed when one plots entanglement entropy vs. disorder strength for this filling, i.e. see figure 6 of ref[17].

From the computational/theoretical viewpoint, figures 1,2 and 3 suggest a number of interesting questions. In particular, why does the entropy method work for filling $^5/_2$ while it is harder to apply to filling $^1/_3$ where the eigenvalue technique works well? This may be due to the observation that a "good" finite size calculation needs two things to happen, firstly, the size of the box should be greater than the physical length scale (proportional to the inverse gap between the ground states and the excited states). In this respect, filling $^1/_3$ with a bigger gap is better. However, since periodic boundary conditions are being used, one also needs the single particle orbitals to overlap strongly, hence a small gap, i.e. filling $^5/_2$ makes it easier numerically to accomplish this.

In any case from an applied and more physical perspective, an issue of greater importance is the role of long range impurities. Experiment indicates (fig. [3] of [5]) that the existence of the $^5/_2$ quantum Hall state is very sensitive to long range impurities. Can this behavior be captured in a relatively simple computational model?

## III. LONG RANGE DISORDER AT FILLING $^5/_2$

Apart from the computational approach taken in the present work, there are several very interesting proposals regarding the strong effect of long range impurities on the $^5/_2$ state[18]. In particular, the $^5/_2$ state can be regarded as a p-wave superconductor of composite fermions and in the context of superconductivity[19] long range impurities can suppress the superconducting order. A second possible scenario is due to the large size of quasi particles in the $5/2$ state; hence the $5/2$ is relatively unaffected by short range impurities, but long range impurities are more dangerous due to the separation of quasi particles into disconnected "puddles"[20].

To study or distinguish either of these pictures computationally will require very large system sizes. This section has the more modest goal of seeing how close a small system size calculation can get to the critical value of the mobility for long range impurities for the $5/2$ state. An estimate of this number is $1.0\ X\ 10^7 \frac{cm^2}{Vs}$ [5,16]compared to a value of $\mu_c$ of $2.0\ X\ 10^6 \frac{cm^2}{Vs}$ [5]for short range disorder.

The model of long range impurities consists of a layer of Coulomb impurities separated from the electron (transport) layer. The model of the layer of Coulomb impurities, is taken from[21]. It is a classical lattice model where the position of Coulomb impurities, assumed to be singly charged and long ranged is obtained through energy minimization using the method described in[21]. In addition to the long ranged potential between impurities, there is an onsite random energy $\Phi_i$ which is taken to be uniformly distributed between $\pm\frac{1}{2}\ b$. The plane of impurities is taken to be $3l_B$ ($l_B$ being the magnetic length) from the electron layer where $l_B$ of[5] is approximately 10 nm. The distance of $3l_B$ , is in contrast to the experimental situation[22] where there are 2 impurity planes placed 75 nm above and below the quantum well of electrons which has a width of 30 nm. The model is therefore a rather coarse-grained description of the more complex experiment of[22]. Since the system sizes treated are of order 20 states in a single Landau level, the box sizes are of order 10 $l_B$ and hence to keep the offset distance smaller than the system size the offset distance was set to $3l_B$.

The effect of the impurity plane on the electron layer is given for a single impurity in[1] p 237, see also[23]. The one impurity result can readily be generalized to many impurities yielding a two-dimensional Fourier transform, working with periodic boundary conditions with a square $a \; x \; a$ cell,

$$V(\vec{q}) = \sum_j \frac{2\pi e^2}{\varepsilon\, q}\, e^{-qC}\, e^{-i\,\vec{q}\cdot\vec{R}_j}\,, (3)$$

where C is the distance of the impurity plane to the electron layer

$$\vec{q} = (\frac{2\pi\, s}{a}, \frac{2\pi\, t}{a}), (4)$$

and $\vec{R}_j$ are the 2 dimensional positions of the impurities.

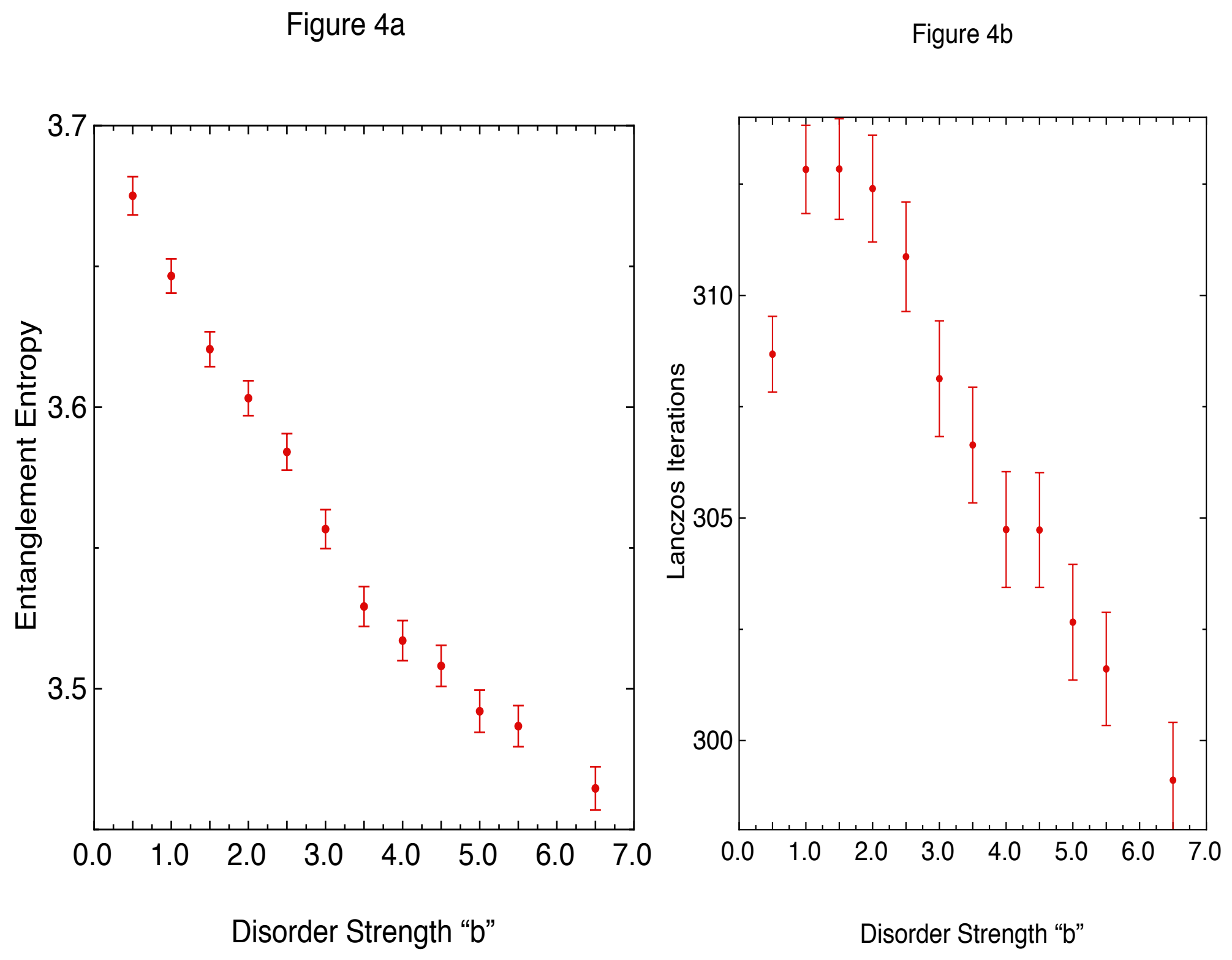

**FIG. 4. Long range impurities for 10 electrons in 20 orbitals filling $^5/_2$. Figure 4a is a plot of entanglement entropy vs. disorder strength where disorder strength refers to "b", the strength of the onsite potential in the impurity plane. Figure 4b is a plot of the average number of initial conditions to converge to the ground state vs. disorder strength "b".**

Due to success using the entanglement entropy to determine the critical mobility for short range impurities for the $5/2$ state, similar calculations have been undertaken for long range impurities. In figure 4a entanglement entropy vs. disorder strength, where in this case disorder strength refers to $b$, the strength of the onsite potential in the impurity plane. At least for the entanglement entropy there is no apparent signature of a transition. Note however, even for short range disorder for this system size[8,17] a transition is not readily apparent.

If however, in figure 4b we plot, starting from the same initial vector, the average number of Lanczos iterations to converge to the ground state, we note a peak near $b = 1.5$. A maximum in this quantity is analogous to a minimum in the gap between the third and fourth excited states for filling $1/3$ in that small gaps suggest a large number of Lanczos iterations.

Figure 5a

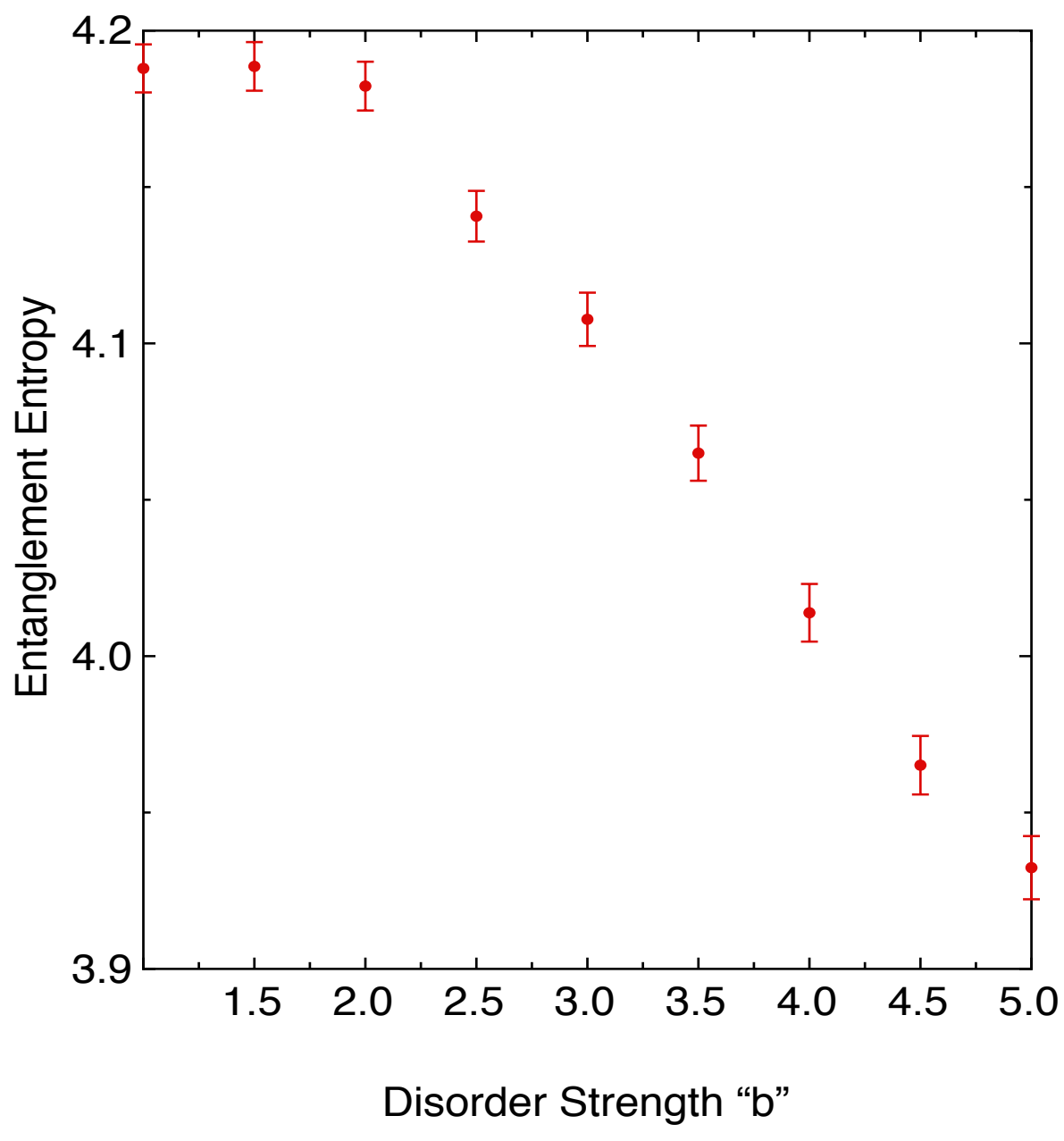

Figure 5b

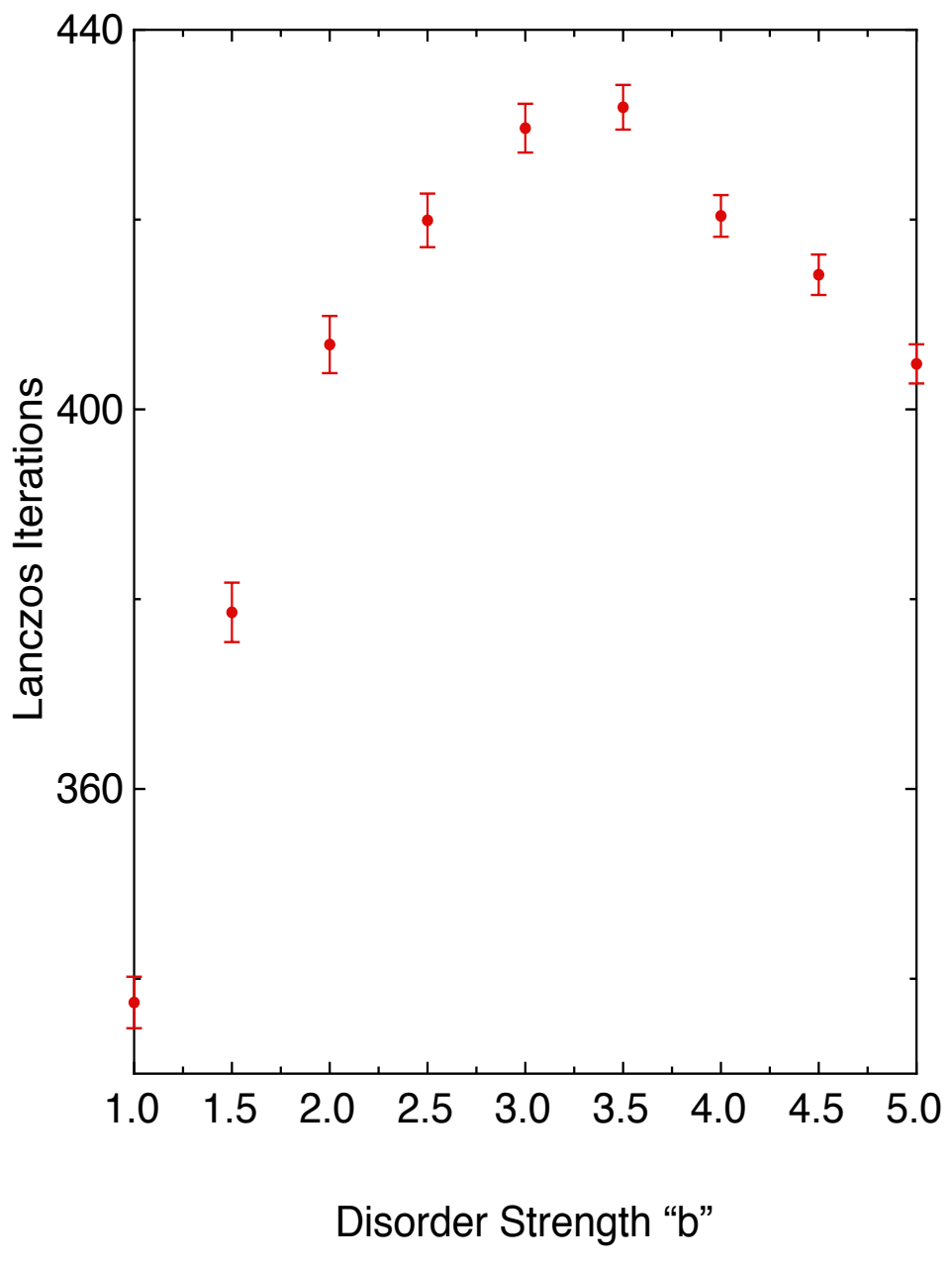

**FIG. 5. Long range impurities for 11 electrons in 22 orbitals filling $^5/_2$. Figure 5a is a plot of entanglement entropy vs. disorder strength where disorder strength refers to "b", the strength of the onsite potential in the impurity plane. Figure 5b is a plot of the average number of initial conditions to converge to the ground state vs. disorder strength "b".**

In the case of short range disorder, a system of 11 electrons in 22 orbitals began to show a clearer transition in entropy vs. disorder strength (see fig. 8b of[8] and figure 7[17]). Therefore, in figure 5a we plot entanglement entropy vs. b for 11 electrons and 22 orbitals (second Landau level) for long range impurities. There appears to be a plateau in the entanglement entropy ending at $b \approx 2$ . Figure 5b shows a peak for $b \approx 3 - 3.5$ in the number of Lanczos iterations for 11 electrons in 22 orbitals. The above figures suggest that these methods could be useful if several more system sizes could be calculated.

Assuming there is a transition at $b \approx 2 - 3$ , what is the corresponding value of the critical mobility $\mu_c$ ? Following the approach of Davies[24] Ch. 8 one finds for b=2 (b=3) mobility of $3.4\,X\,10^7 \frac{cm^2}{Vs}$ ($2.5\,X\,10^7 \frac{cm^2}{Vs}$ ). These values are three times larger than the critical mobility for long range impurities found in experiment. Whether this is due to unrealistic modelling or finite size effects is a subject of future study.

## IV. CONCLUSIONS

Short and long-range impurities have been examined for fractional quantum Hall systems. There appears to be a consistent computational picture for short range impurities. In the case of long range impurities, calculations agree qualitatively with experiment, in that the critical mobility is very sensitive to long range impurities and the critical mobility for long range impurities is larger than the critical mobility for short range impurities. The physical mechanism of this sensitivity and a quantitative understanding remain a challenging computational issue. A relatively straight forward effort could treat four and possibly five more system sizes (12-24 through 16-32) and handle two rather than one impurity planes. However, for 32 orbitals the box dimension $a$ is still a "mere" 14.1 $l_B$. To treat larger systems, say 60 orbitals, the density matrix renormalization group (dmrg)is an interesting possibility. However, treating disorder may not be straight forward (though see[25]) and calculating the entanglement entropy accurately can be challenging within the dmrg approach[8].

After finishing a draft of the present paper, another interesting relevant paper was published[26]. This paper,[26] extends the study of [11] to include long range Coulomb interaction between electrons. The value of $W_c$ is thus directly comparable to[6,7] and to the value obtained here. In[6,7] for long range Coulomb potential and filling $1/3$, a value of $W_c \approx .20$ compared to $W_c \approx .15$ calculated in the present paper and $W_c \approx .09$ of[26]. By using the Born Approximation result for the mobility[6,7] the recent paper[26] gives a critical mobility roughly 2.5 times the experimental value. Note that the techniques to obtain the $W_c$ in the present work and[11,26] are quite distinct. As well as a more detailed comparison of the numerical methods, it would be of interest to have an additional systematic experiments studying the critical mobility for fractional quantum Hall systems in the lowest Landau level.